\documentclass[prc,aps,superscriptaddress,showpacs,nofootinbib,twocolumn]{revtex4}
\usepackage{amssymb}
\usepackage[dvips]{graphicx}
\usepackage[english]{babel}
\usepackage{indentfirst}
\usepackage{amsxtra}
\usepackage{amsmath}
\usepackage{supertabular}
\usepackage{multirow}
\usepackage[mathcal]{eucal}
\usepackage[usenames]{color}
\usepackage{ulem}

\begin{document}
\title {\textbf{Beyond--mean--field effective masses in the nuclear Fermi liquid from axial breathing modes}}
\author{M. Grasso}
\affiliation{Institut de Physique Nucl\'eaire, CNRS-IN2P3, Universit\'e Paris-Sud,
Universit\'e Paris-Saclay, 91406 Orsay, France}
\author{D. Gambacurta}
\affiliation{Extreme Light Infrastructure - Nuclear Physics (ELI-NP), Horia Hulubei National Institute for Physics and Nuclear
Engineering, 30 Reactorului Street, RO-077125 M˘agurele, Jud. Ilfov, Romania}
\author{O. Vasseur}
\affiliation{Institut de Physique Nucl\'eaire, CNRS-IN2P3, Universit\'e Paris-Sud,
Universit\'e Paris-Saclay, 91406 Orsay, France}

\begin{abstract}
Axial breathing modes are studied within the nuclear energy--density--functional theory to discuss  the modification of the nucleon effective mass produced beyond the mean--field approximation. This analysis is peformed with the subtracted second random--phase--approximation (SSRPA) model applied to two nuclei, $^{48}$Ca and $^{90}$Zr. Analyzing the centroid energies of axial breathing modes obtained with the mean--field--based random--phase approximation and with the beyond--mean--field SSRPA model, we estimate the modification (enhancement) of the effective mass which is induced beyond the mean field. This is done by employing a relation, obtained with the Landau's Fermi liquid theory, between the excitation frequency of axial modes to $\sqrt{m/m^*}$, where $m$ ($m^*$) is the bare (effective) mass. Such an enhancement of the effective mass is discussed in connection with the renormalization of single--particle excitation energies generated by the energy--dependent SSRPA self-energy correction. We find that the effective beyond--mean--field compression of the single--particle spectrum produced by the self--energy correction is coherent with the increase of the effective mass estimated from the analysis of axial breathing modes.  
\end{abstract}

\maketitle
% {\color{red} prova }
%\section{Introduction}

The Landau's theory of interacting Fermi systems \cite{pines,abri} provides an elegant description of 
low--energy excitations in Fermi liquids, for different types of many--body systems and at different energy scales. The complex dynamics of the interacting particles is simplified through the concept of 
 quasiparticles having an effective mass $m^*$ induced by the interparticle interaction.
The study of $m^*$ meets a broad interest in several branches of many--body physics. Being $m^*$ related to the propagation of particles in a medium and, more specifically, to the density of states in many--body systems \cite{fetter}, it has an important impact on several observables such as, for instance, the energies of axial compression modes in atomic gases \cite{nasci} and in nuclei \cite{blaizot1,bohigas}, the specific heat of a low--temperature Fermi gas \cite{fetter}, or the 
maximum mass of a neutron star (the maximum mass that an equation of state may generate taking into account gravity) \cite{gle}.

The effective mass is usually evaluated through the computation of the self--energy (see for instance Refs. \cite{hedin,fetter}). Numerically, it may be provided by Quantum Monte Carlo calculations, for instance as done in Refs. \cite{lobo,pilati} for the polaron in strongly imbalanced atomic gases. 
Analogous calculations were also done in Refs. \cite{forbes,roggero} for nuclear systems.
It was also proposed in Ref. \cite{eich} to extract the effective mass from thermodynamical properties in an interacting electron liquid. 

For ultracold atomic Fermi gases, the properties of the polaron quasiparticle have been extensively studied \cite{na1,sc,nasci,sca,koh,kos}. In the case of   
 very large population imbalance, the measurement of low--frequency axial breathing modes has been  used to extract dynamically the polaron effective mass \cite{nasci}: based on the Landau theory of Fermi liquids, a relation was drawn in the local--density approximation between the frequency of the polaron $\omega^*$ and its effective mass $m^*$, such that $\omega^*$ is proportional to $\sqrt{m/m^*}$, $\omega^*/\omega = \sqrt{(1-A)m/m^*}$, where $\omega$ is the trapping--potential frequency and $A$ describes the attraction between the impurity and the other atoms. An analogous relation, based on the Landau theory of Fermi liquids, was employed several decades ago for atomic nuclei \cite{blaizot1,bohigas} to connect the centroid energies of isoscalar (IS) giant quadrupole resonances (GQRs) (which are the nuclear axial breathing modes)  with the effective mass in nuclear matter. The local--density approximation can be applied in this case because the effective mass is a smooth function of the density. Such a relation was widely used in nuclear physics in the past decades to extract, from the measurement of IS GQRs, phenomenological constraints for the effective mass in matter (see Ref. \cite{bao} and references therein).  

The effective mass $m^*$ is defined by the relation 
\begin{equation}
\frac{1}{m^*}=\frac{dE}{dk} \frac{1}{\hbar^2k} \,
\label{defimstar}
\end{equation}
for a particle of energy $E$ and momentum $k$, with
\begin{equation}
E=\frac{\hbar^2 k^2}{2m} + \Sigma_k + \Sigma_{k,E}. 
\label{ener}
\end{equation}
In Eq. (\ref{ener}), $ \Sigma_k + \Sigma_{k,E}$ is the self--energy, sum of the mean--field (MF) contribution $\Sigma_k$ (from the leading order of the Dyson equation in the perturbative many--body expansion) and of a beyond--mean--field (BMF) energy--dependent contribution $\Sigma_{k,E}$. 
In the MF approximation, the self--energy does not have any energy 
dependence and may have only a $k$ dependence. An explicit energy dependence is induced when the MF approximation is overcome going beyond the leading order of the Dyson expansion \cite{fetter}. 
Using the definition of $m^*$ in Eq. (\ref{defimstar}), one can write
\begin{eqnarray}
\nonumber
\frac{m^*}{m}&=&\left(1-\frac{\partial \Sigma_{k,E}}{\partial E}\right) \cdot \left(1+\frac{m}{\hbar^2k}\frac{\partial \Sigma_k}{\partial k}\right)^{-1} \\ 
&=& \frac{m_E^*}{m} \cdot \frac{m_k^*}{m}, 
\label{mstar}
\end{eqnarray}
where the above expression defines the so--called $E$--mass $m_E^*/m$ and $k$--mass $m_k^*/m$, using the same notations as in Refs. \cite{blaizot1,jeu,ma,bla,bern}. 
In cases where the MF self-energy does not have a $k$ dependence (for instance, with a zero--range interaction characterized only by a coupling constant, without any velocity--dependent terms) the $k$--mass is equal to 1. 
In these cases, an effective mass is generated only with second--order calculations.

\begin{figure}
\includegraphics[scale=0.35]{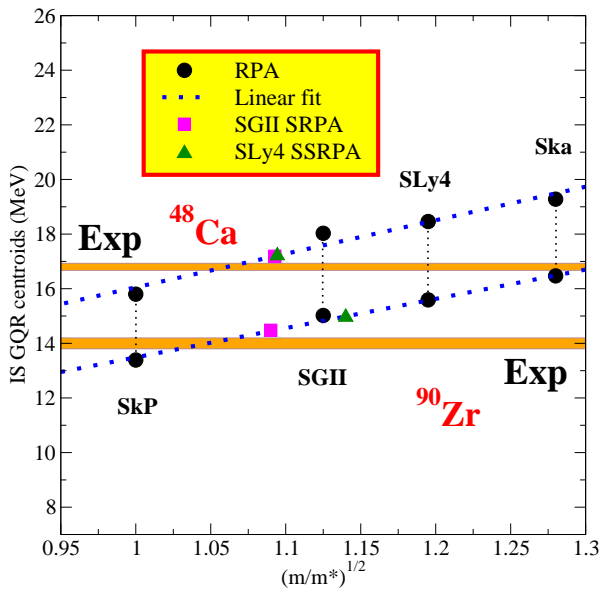}
\caption{Centroid energies of the IS GQRs for the nuclei $^{48}$Ca and $^{90}$Zr as a function of $\sqrt{m/m^*}$. The RPA centroids (black circles) are reported for four Skyrme parametrizations and associated to the corresponding MF effective masses in nuclear matter. A linear fit is done on these points (blue dotted lines). The SSRPA-SLy4 and SSRPA-SGII centroids are reported on the blue dotted lines (green triangles and magenta squares, respectively). The experimental values are also displayed by orange bands.}
\label{effem}
\end{figure}

The $E$-mass is equal to 1 in the MF approximation, where $m^*=m^*_k$. 
Any BMF effect produces a modification of $m^*$ generated by the $E$-mass. 
Being the effective mass related to the density of states \cite{erler}, BMF changes of its value induce a different single--particle spectrum, which is compressed if the effective mass is enhanced beyond the mean field. 

This aspect is investigated here with the SSRPA model introduced in Ref. \cite{gamba2015}, where the MF approximation is overcome owing to the coupling of 1 particle-1 hole (1p1h) and 2 particle-2 hole (2p2h) configurations. The study is done in the framework of the energy--density--functional (EDF) theory \cite{bender} with Skyrme forces. 
We base our analysis on the above--mentioned relation between the frequency of axial modes and $\sqrt{m/m^*}$. 
We propose a new and original procedure to estimate BMF effects on the effective mass of nuclear matter. This is based on BMF predictions of axial breathing modes in nuclei and is connected with an induced BMF modification of single--particle spectra. This procedure is quite general and can be employed with other BMF models. Nevertheless, the SSRPA has an important advantage (compared to other BMF models) of being definitely safe against instabilities, divergences, and double counting of correlations  \cite{gamba2015,tse} in the EDF framework. This guarantees the quantitative robustness of the obtained predictions. 

We first performed random--phase--approximation (RPA) calculations for the medium--mass nucleus $^{48}$Ca and the heavier nucleus $^{90}$Zr by using four Skyrme parametrizations SkP \cite{skp}, SGII \cite{sgii}, SLy4 \cite{sly4}, and Ska \cite{ska} having, respectively, MF effective masses equal to 1, 0.79, 0.7, and 0.61 in nuclear matter. 
We plot in Fig. \ref{effem} the obtained centroid energies for the IS GQR modes as a function of $\sqrt{m/m^*}$, associating to each centroid energy the corresponding MF effective mass in matter. A linear fit is performed on these four points for each nucleus (blue dotted lines) and the experimental values are also displayed (orange bands). 
 
We choose two parametrizations, SLy4 and SGII, having MF effective masses between 0.7 and 0.8.
To estimate the modification of the effective mass produced beyond the mean field we use the linear fits performed on the RPA points and we report, on the blue dotted lines, the points corresponding to the SSRPA centroid energies obtained for the two nuclei and the two parametrizations. We observe that the centroids are located at lower energies for the SSRPA model with respect to the corresponding RPA values. Such a lowering of the energies implies that the associated effective mass increases with respect to the MF value. 

We deduce that, 
 for $^{48}$Ca ($^{90}$Zr), the extracted effective mass for nuclear matter increases from 0.7 in the MF case to 0.834 (0.769) for the BMF  calculations of the IS GQR with SLy4. With SGII, the  effective mass for matter increases from 0.79 to 0.837 (0.842) from the calculations done for $^{48}$Ca ($^{90}$Zr). 

\begin{figure}
\includegraphics[scale=0.34]{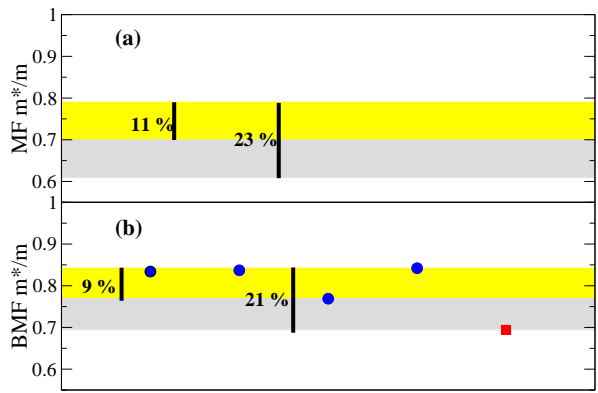}
\caption{(a) Theoretical error associated to the MF effective mass for nuclear matter induced by two Skyrme parametrizations, SLy4 and SGII (yellow band) and three Skyrme parametrizations, SLy4, SGII, and Ska (yellow plus grey band); (b) Same as in panel (a) but for the BMF effective mass. The four blue circles represent the calculations done with SLy4 and SGII for $^{48}$Ca and $^{90}$Zr whereas the red square represents the calculation done with Ska for $^{48}$Ca.}
\label{window}
\end{figure}

Figure \ref{window} displays an estimation of the theoretical error bar associated to the spreading of the values of the effective mass in matter. Figure \ref{window}(a) shows the MF error bar (yellow area) which is induced by the dependence on the used interaction of the nuclear matter $m^*$ value (11\% of discrepancy between the SLy4 and SGII values). Figure \ref{window}(b) shows the four BMF values for the effective mass (blue circles). One notices that the maximum discrepancy is now of 9\% (yellow area), slightly reduced with respect to the MF case even if, in the 
BMF case, the extracted $m^*$ value depends not only on the used interaction but also on the nucleus for which the axial excitation is computed. The two nuclei under study already offer the possibility to cover two different mass regions (nucleus dependence). To 
extend our analysis we performed an additional SSRPA calculation for $^{48}$Ca (the nucleus for which we have found the largest modification of effective mass going from MF to BMF) with the parametrization Ska (MF $m^*=0.61$). The MF theoretical error increases correspondingly (maximum discrepancy of 23\%, yellow + grey area in Fig. \ref{window}(a)). We observe that, including the Ska BMF value in Fig. \ref{window}(b) (red square) the discrepancy window is now represented by the yellow plus grey area (21 \% of maximum discrepancy). We may thus deduce that such an extraction of a BMF effective mass for nuclear 
matter does not produce an overall error larger than the one already induced with MF calculations and related to the dependence on the used interaction.

From Eq. (\ref{mstar}), we may extract the average values of the $E$-mass, equal to 1.19 (1.06) with SLy4 (SGII) for $^{48}$Ca and 
to 1.10 (1.07) with SLy4 (SGII) for $^{90}$Zr. The $E$-mass is equal to 1.14 for $^{48}$Ca with Ska. BMF effects produce an increase of the $E$-mass ranging from 6 to 16 \%, the largest variation from 1 occurring for $^{48}$Ca and the SLy4 parametrization.

%\begin {table} 
%\begin{center}
%\begin{tabular}{cccccc}
%
%                     \hline
%\hline
% Parametrization                     & $^{48}$Ca & $^{48}$Ca &  & $^{90}$Zr & $^{90}$Zr\\
%         &      $m^*/m$ & $E$-mass & & $m^*/m$ & $E$-mass \\
%\hline
%  SLy4                   & 0.834 & 1.19 & & 0.769 & 1.10       \\
%   SGII & 0.837 & 1.06     &  & 0.842 & 1.07 \\
%\hline
%\hline
%\end{tabular}
%\end{center}
%\caption{Average values of the effective mass and of the $E$-mass for the nuclei $^{48}$Ca and $^{90}$Zr and the Skyrme parametrizations SLy4 and SGII.}
%\label{emass}
%\end {table}  

We observe in Fig \ref{effem} that, for $^{48}$Ca, the SSRPA centroid energies obtained with the two parametrizations SLy4 and SGII are very similar, leading to very similar values of $m^*/m$. Since the MF effective mass is not the same for the two parametrizations, this implies a stronger BMF modification of the 
$E$-mass for the case of SLy4 (where the MF effective mass is lower). On the other side, the SSRPA centroids obtained with SLy4 and SGII are slightly different for $^{90}$Zr, leading to a higher value of $m^*/m$ for the case of SGII. The two BMF $E$-masses are very similar one to the other for this nucleus. 

It is interesting to mention that the lowering 
of the excitation energies provided by SRPA--based models (with respect to the RPA spectrum) is a general  
feature of the model that does not occur only in nuclear systems. The same type of effect was found for instance also for metallic clusters in Ref. \cite{gamba2009,gamba2010}. In all cases, one thus expects an increase of the 
effective mass ($E$-mass larger than 1) and, consequently, an effective compression of the single--particle spectrum. 

We analyze this aspect in the framework of the BMF SSRPA model. 
Writing the SRPA equations as energy--dependent RPA equations, the RPA--type energy--dependent matrix elements contain the RPA matrix elements plus additional contributions given by the energy--dependent self--energy (see for instance Ref. \cite{papa}). For example, the energy--dependent $A$-type matrix elements may be written as 
\begin{equation}
A^{SRPA}_{1,1'} (E) = A^{RPA}_{1,1'}+\sum_{2,2'} \frac{A_{12} A_{2'1'}}{E+i\eta-A_{2,2'}},
\label{asrpa}
\end{equation}
where $A^{RPA}_{1,1'}$ are standard RPA $A$ matrix elements, 1 and 1' are two 1p1h configurations, 2 and 2' are two 2p2h configurations, and $A_{12}$ and $A_{22}$ are beyond--RPA matrix elements coupling 1p1h with 2p2h configurations and 2p2h configurations among themselves, respectively. 
For the sake of simplicity, we write the expressions for cases where the interaction is density independent, where rearrangement terms are not present. The expressions which are used in practice for the results presented in Fig. \ref{effem} are slightly more involved and contain additional contributions corresponding to the rearrangement terms \cite{rearra}. 

%\begin{figure}
%\includegraphics[scale=0.35]{ener48Casly4.eps}
%\caption{Diagonal matrix elements $A_{1,1}$ calculated with the parametrization SLy4 for the nucleus $^{48}$Ca for the first three single--particle configurations (which are neutron configurations). RPA, SRPA, and SSRPA results are presented. The BMF results are calculated using in the energy--dependent matrix elements an energy value equal to the average between the two IS GQR centroids obtained in SRPA and in SSRPA.}
%\label{ene48}
%\end{figure}

%\begin{figure}
%\includegraphics[scale=0.35]{ener90Zrsly4.eps}
%\caption{Same as in Fig. \ref{ene48} but for $^{90}$Zr. One of the configurations is in this case a proton configuration.}
%\label{ene90}
%\end{figure}

The energy--dependent self--energy correction provides a renormalization of the diagonal matrix elements 
$A_{1,1}$ (corrections to both the single--particle excitation energies and the interaction matrix elements). Since such matrix elements contain the single--particle excitation energies, this renormalization certainly induces a BMF renormalization of the 1p1h single--particle spectrum. One has 
\begin{eqnarray}
\nonumber
&& A_{1,1}^{RPA} =
\left[\epsilon_p - \epsilon_h \right]_{MF} + \bar{V}_{phhp} \rightarrow  A_{1,1}^{SRPA} (E) \\
&=&\left[\epsilon_p - \epsilon_h\right]_{MF} + \bar{V}_{phhp} + \sum_{2,2'} \frac{A_{ph,2} A_{2',ph}}{E+i\eta-A_{2,2'}}.
\label{reno}
\end{eqnarray}
The subtraction procedure used in Ref. \cite{gamba2015} was formulated in Ref. \cite{tse} so to avoid any double counting of correlations in extensions of RPA within the framework of EDF theories. 
The subtraction procedure is based on the dielectric theorem \cite{bohigas}
stating that
the inverse energy--weighted moment of the strength evaluated in RPA is
equal to the static polarizability. In the SRPA, because of the energy
dependence introduced into the self--energy, this
equality is not fulfilled anymore. The subtraction procedure restablishes
the equality of the RPA and SSRPA inverse energy--weighted moments and, as
a consequence,  the dielectric theorem.
This is achieved by imposing that the beyond--RPA energy--dependent matrix elements are equal to the corresponding RPA matrix elements in the static limit (self--energy calculated at zero energy). For details, the reader may refer to Refs. \cite{gamba2015,tse}.

\begin{widetext}

\begin{figure}
\includegraphics[scale=0.45]{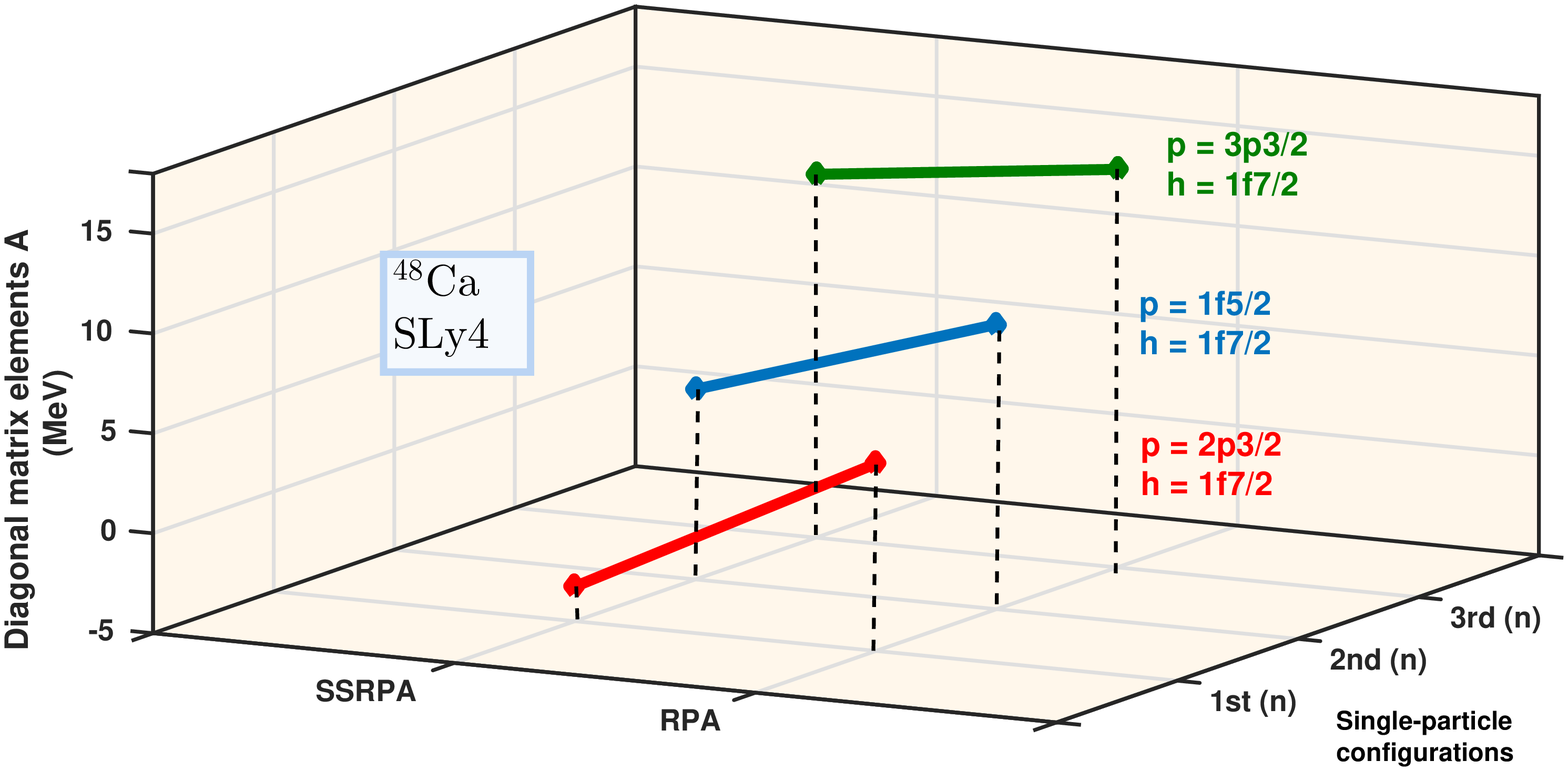}
\caption{Diagonal matrix elements $A_{1,1}$ calculated with the parametrization SLy4 for the nucleus $^{48}$Ca for the first three single--particle configurations (which are neutron configurations). RPA and SSRPA results are presented. The BMF results are calculated using in the energy--dependent matrix elements an energy value equal to the IS GQR centroid obtained in SSRPA.}
\label{ene48}
\end{figure}

\begin{figure}
\includegraphics[scale=0.45]{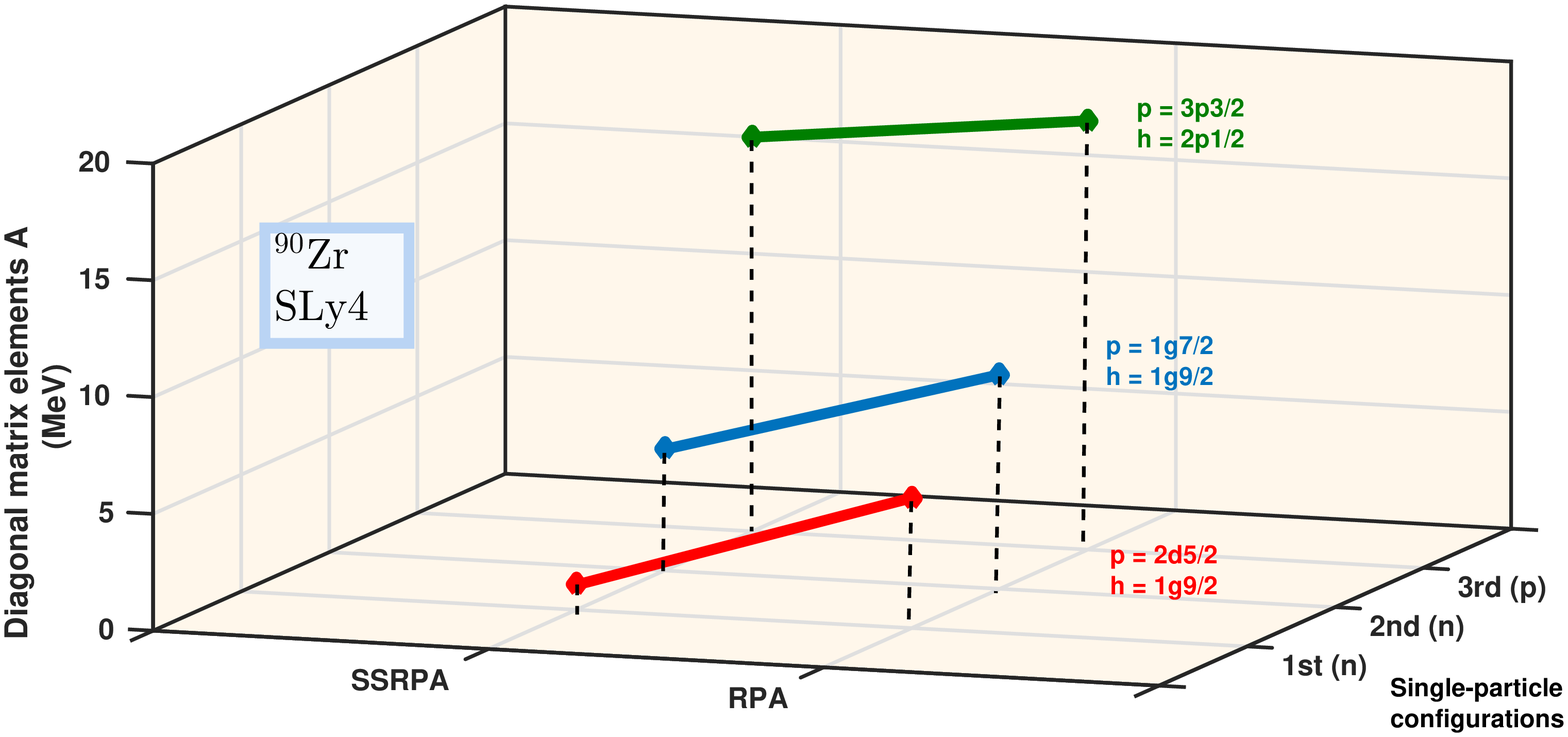}
\caption{Same as in Fig. \ref{ene48} but for $^{90}$Zr. One of the configurations is in this case a proton configuration.}
\label{ene90}
\end{figure}

\end{widetext}

By performing the subtraction procedure, the rescaling of the matrix element $A_{1,1}$  is thus further modified by an additional corrective term, which guarantees that $A^{SSRPA}_{1,1}(0)=A^{RPA}_{1,1}$,
\begin{eqnarray}
\nonumber
&& A_{1,1}^{RPA}  \rightarrow A_{1,1}^{SSRPA} (E) =  \left[\epsilon_p - \epsilon_h \right]_{MF}  
+ \bar{V}_{phhp}  \\ &+& \sum_{2,2'} \frac{A_{ph,2} A_{2',ph}}{E+i\eta-A_{2,2'}}  
+ \sum_{2,2'} \frac{A_{ph,2} A_{2',ph}}{A_{2,2'}}.
\label{renosub}
\end{eqnarray}
As an illustration, we discuss the case of the parametrization SLy4. 
Figure \ref{ene48} shows the diagonal matrix elements $A_{1,1}$ calculated for the nucleus $^{48}$Ca for the first three 1p1h configurations entering in the construction of the collective quadrupole excitations. In this case, the three configurations are neutron configurations. We present RPA and SSRPA results. In the case of  SSRPA, to compute the rescaling effect induced by BMF calculations, the energy--dependent self--energy correction is calculated at an energy value given by the SSRPA centroid of the IS GQR. This guarantees that we make this estimation in the region of the GQR. 

We notice that the BMF rescaling of the matrix element is more pronounced for the first configuration and becomes less important at increasing energies. We have observed that this effect is indeed strongly quenched for the highest--energy 1p1h configurations. 
Also, the BMF modification produces in all cases a global reduction of the matrix element (and, consequently, a reduction of the single--particle excitation energy). Such a reduction implies an effective compression of the single--particle spectrum, coherent with the enhancement of the effective mass indicated by our previous analysis done on the values of the centroid energies. 

Figure \ref{ene90} shows the same quantities as Fig. \ref{ene48}, but for the nucleus $^{90}$Zr. In this case, the third single--particle configuration is a proton configuration. 

For the two nuclei, $^{48}$Ca and $^{90}$Zr, the third single--particle configuration entering in the construction of the quadrupole collective phonon is located in the same energy region as the IS GQR. One can thus in this case provide an intuitive physical interpretation of the BMF renormalization effects. Such effects  may be regarded in the same spirit as in particle--phonon--coupling models: the unperturbed 1p1h excitation mode couples with the collective phonon because the two excitation energies are close one to the other. The resulting effect is the lowering of the centroid for the collective phonon, the formation of a spreading width for the collective excitation due to the mixing with 2p2h configurations, and the compression of the single--particle spectrum that we deduce from the fact that our effective single--particle excitation energies are systematically reduced. 
Let us focus on this third 1p1h configuration and on the nucleus $^{48}$Ca. 
Figure \ref{ene48} shows that the SSRPA model induces a reduction of 11.6 \% of the matrix element $A_{1,1}$ with respect to the RPA result. 

\begin{figure}
\includegraphics[scale=0.35]{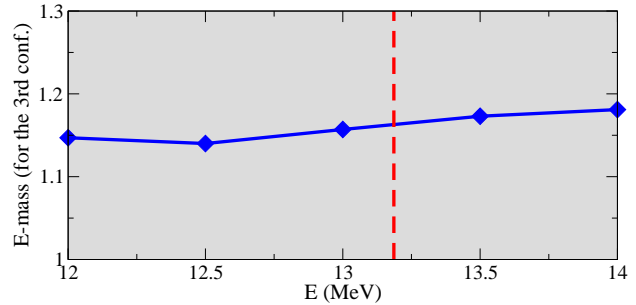}
\caption{Estimation of the $E$-mass (for the third single--particle configuration) for the nucleus $^{48}$Ca and the parametrization SLy4. The red dashed line represents the value of the diagonal matrix element $A_{1,1}$ for the third 1p1h configuration. }
\label{sigma}
\end{figure}

To push further this analysis, we compute the SSRPA energy--dependent self--energy correction using the second line of Eq. (\ref{renosub}). We have done this calculation for the third single--particle configuration,  for the nucleus $^{48}$Ca and the parametrization SLy4. We denote this self--energy by $\Sigma_{3}$. We show in Fig. \ref{sigma}, as a function of the energy, the quantity $1-\partial \Sigma_3/\partial E$, which should  
 correspond to an estimation of the $E$-mass for the third 1p1h configuration (see Eq. (\ref{mstar})). The derivative of $\Sigma_3$  with respect to the energy is negative, which leads to values for the $E$-mass  larger than 1. 
This quantity goes to 1 in the static limit where we 
recover the MF result. Coherently with the previous extraction of the average $E$-mass from the centroid of the collective axial modes, we also notice that, in the energy region of the third 1p1h configuration (the red dashed line in the figure represents the value of the diagonal matrix element $A_{1,1}$ for the third 1p1h configuration), for which the computation is performed, this estimation of the $E$-mass provides the value of 1.16, which is very close to the one previusly found for $^{48}$Ca and SLy4. 

Analogous results were obtained for the heavier nucleus $^{90}$Zr and the parametrization SGII. In general, we did not identify any particular difference between the results obtained for the two nuclei. Their different mass does not lead to any specific dissimilarities in the renormalization of the effective mass and of the single--particle spectrum. 

In conclusion, we have discussed the enhancement of the nucleon effective
mass induced by BMF effects. The study is based on a new procedure that was applied within the EDF theory. The SSRPA model is employed, which allows for a
microscopic description of BMF effects. The average BMF effective mass is extracted from the
predictions of the centroid energies of axial breathing modes in two
nuclei, $^{48}$Ca and $^{90}$Zr. The compression of single--particle spectra
generated by the SSRPA self--energy correction is also investigated. The
two analyses, one based on the centroid energies of axial modes and the other based on the compression of
single--particle spectra, lead to a coherent estimation of the average $E$-mass value,
which increases from 6 to 16 \%, depending on the nucleus and on the effective interaction, with respect to the MF value, which is equal to 1.

%\begin {table} 
%\begin{center}
%\begin{tabular}{cccccc}
%
%                     \hline
%\hline
%                     & Exp           & SRPA & SSRPA & SRPA & SSRPA\\
%  &   &     SGII         &  SGII &   SLy4 & SLy4  \\
%\hline
%$\sum B(E1)$         & 0.068       & 0.563 & 0.078 & 1.012& 0.126\\
%                     & $\pm$ 0.008 &       &       &      &       \\
%\hline
%$\sum_i E_i B_i(E1)$ & 0.570       & 4.618& 0.621 & 8.795& 1.062\\
%                     & $\pm$ 0.062 &       &       &      &       \\
%\hline
%\hline
%\end{tabular}
%\end{center}
%\caption{ Experimental and theoretical $\sum B(E1)$ in ( e$^{2}$ fm$^{2}$) and $\sum_i E_i B_i(E1)$ in (MeV e$^{2}$ fm$^{2}$) 
%summed between 5 and 10 MeV.}
%\end {table} 

%
%
%
%-----------------------------------

\end{document}